\documentclass[10pt,aps,twocolumn,preprintnumbers,amsmath,amssymb,floatfix]{revtex4-1}
\usepackage{graphicx}

\begin{document}
\title{Detection of adsorbed transition-metal porphyrins by spin-dependent conductance of graphene nanoribbon}
\author{Peter Kratzer}
\affiliation{School of Physics, The University of Sydney, Sydney, New South Wales 2006, Australia}
\affiliation{Faculty of Physics, University of Duisburg-Essen, 47057 Duisburg, Germany}
\author{Sherif Abdulkader Tawfik$^*$}
\affiliation{School of Physics, The University of Sydney, Sydney, New South Wales 2006, Australia}
\affiliation{$^*$present address: School of Mathematical and Physical Sciences, University of Technology Sydney, Ultimo, New South Wales 2007, Australia}
\author{X. Y. Cui}
\affiliation{Australian Centre for Microscopy and Microanalysis, and School of Aerospace,
Mechanical and Mechatronic Engineering, The University of Sydney, Sydney, New South Wales 2006, Australia}
\author{C. Stampfl}
\affiliation{School of Physics, The University of Sydney, Sydney, New South Wales 2006, Australia}\date{\today}
\begin{abstract}
Electronic transport in a zig-zag-edge graphene nanoribbon (GNR) and its modification by adsorbed transition metal porphyrins is studied 
by means of density functional theory calculations.  The detachment reaction of the metal centre of the porphyrin is investigated both in the gas phase and for molecules adsorbed on the GNR. As most metal porphyrins are very stable against this reaction, it is found that these molecules bind only weakly to a {\em perfect} nanoribbon. However, interaction with a single-atom vacancy in the GNR results in chemical bonding by  the transition metal centre being shared between nitrogen atoms in the porphyrin ring and  the carbon atoms next to the vacancy in the GNR. For both the physisorbed and the chemisorbed geometry, the inclusion of van der Waals interaction results in a significant enlargement of the binding energy and reduction of the adsorption height. Electronic transport calculations using non-equilibrium Greens functions show that the conductivity of the GNR is altered by the chemisorbed porphyrin molecules. Since the metal centers of porphyrins carry an element-specific magnetic moment, not only the net conductance, but also the spin-dependent conductance of the GNR is affected. In particular, the adsorption of Ru-porphyrin on the single-atom vacancy results in a very large spin polarization of the current of 88\% at small applied source-drain voltages. Based on our results, we suggest that a spin valve constructed from a GNR with ferromagnetic contacts could be used as a sensitive detector that could discriminate between various metal porphyrins. 
\end{abstract}

\maketitle
\section{Introduction}

Metal porphyrins constitute a class of versatile molecules that play an important role in diverse branches of science, such as biochemistry and materials science. 
For instance, the heme unit in the hemoglobin molecule, responsible for the red colour of blood, is an iron porphyrin. The chlorophyll of green plants contains Mg porphyrins. Moreover, nature uses various other metal porphyrins as biocatalytic units in enzymes. In materials science, metal porphyrins \cite{Higashino2015},  metal phthalocyanines \cite{Karousis2012} and related organic metal complexes \cite{Kalyanasundaram1998}  
have been considered as light absorbers in solar cells. Furthermore, 
a porphyrin molecule perched between electrodes made of metal clusters or graphene \cite{Rakhmilevitch2014,Zhou2013,Schmaus2011} has been suggested in the context of molecular electronics. 
In the present paper, we focus on a related topic, namely, how electrical transport can be modified by the presence of porphyrin molecules. We suggest a different approach, based on an electrical current flowing through an intact graphene nanoribbon, whose conductance is modified by the adsorption of a metal porphyrin. 
This new geometry, with the porphyrin sitting outside the current path, allows us to combine a high conductance 
with a sufficiently high sensitivity of the current to an adsorption event. Hence we suggest to use this geometry, rather than a molecule clamped between two leads, for a detector for metal porphyrins based on a graphene nano-device. 
The idea of using the transport properties of graphene for ultra-sensitive detection of chemicals on the single-molecule level has already been proposed, e.g. in Ref.~\cite{Tawfik2015}. 
While many metal porphyrins are used as dyes and hence show a clear signal in optical spectroscopy that can be used for their detection, a solid-state detector accessible to simple electrical read-out could have many advantages, such as lower cost and wider applicability.  

We use a zig-zag-edged graphene nanoribbon (GNR) whose C atoms at the edges are saturated by hydrogen atoms. 
These nanoribbons are metallic \cite{Dubois2009} and offer a considerable conductivity due to edge channels. 
Specifically, we study a GNR that is seven C atoms in width. 
An odd number is chosen deliberately in order to avoid the symmetry-selection rule that suppresses conductivity in 
GNRs of even width~\cite{Li2008}. 
Graphene and graphene nanoribbons have attained much interest in recent years also in the context of spintronics. 
Here, the large spin diffusion length in graphene makes this material a well-suited spin conductor. 
The demonstration of ferromagnetic metallic electrodes attached to graphene by van Wees and co-workers~\cite{Tombros2007,Jozsa2008} has opened up the route to spin injection into graphene and related carbon materials, and to a  carbon-based spintronics. 
In the present context, this implies that not only the conductance itself, but also its spin dependence could be used as a fingerprint that could allow us to add chemical sensitivity to a graphene-based detection device. 
Moreover, if the adsorbed porphyrin molecule possesses a magnetic moment, an external magnetic field can be used to manipulate the transport properties \cite{Candini2011}.

\begin{figure}[btp]
\begin{center}
\includegraphics[width=0.48\textwidth]{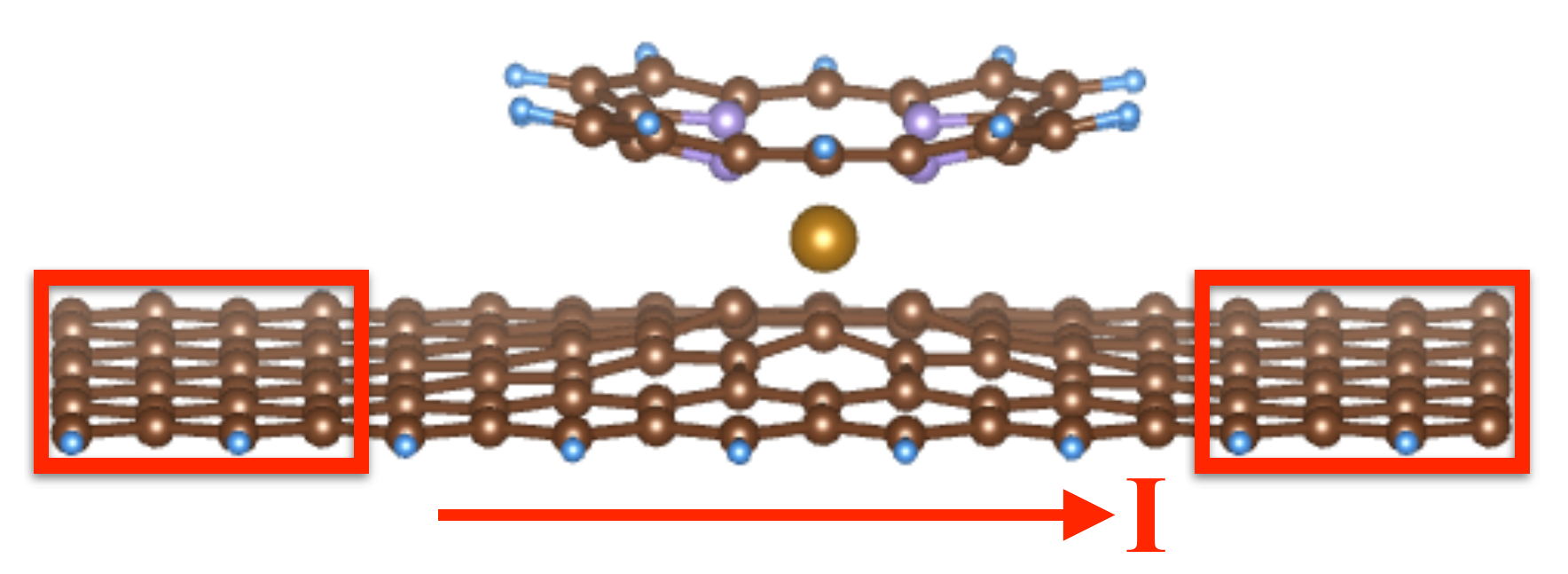}
\caption{Geometrical set-up used for the transport calculations. The graphene nanoribbon with single-atom vacancy is modeled by C$_{125}$H$_{18}$. 
Contact regions (red boxes) with fixed atomic positions are defined at both ends. The inner region without the contacts corresponds to C$_{69}$H$_{10}$. The pictures shows the adsorption geometry of a Fe-porphyrin using the PBE functional. The Fe atom (orange ball) leaves the porphyrin ring and reaches a 'sandwich position' between the nanoribbon and the porphyrin.}
\label{fig:set-up}
\end{center}
\end{figure}

Metal porphyrins are organic ring molecules with a central metal atom. This metal centre  is bonded to nitrogen atoms that are part of pyrrole subunits (5-membered rings), similar to the bonding of transition metal complexes in coordination chemistry. Formally, the metal porphyrins can be derived from a metal-free form, the porphine (sometimes also called the 'free base') in which two hydrogen atoms, bonded to two of the N atoms, replace the metal in the ring centre. Many elements, including both simple as well as transition metals, form stable metal porphyrins. 
Adsorption of organic ring molecules with a metal centre has been studied at low temperature on graphene films, in particular on films  grown on surfaces of transition metals, e.g. Ni \cite{Klar2014} or Ir \cite{Endlich2014}. 
For the envisaged application as a detector, however, we require a strong chemical interaction between the molecules and the GNR which allows us to operate at room temperature and provides us with a mechanism to affect the transport in the GNR. 
For example, chemisorption of a molecule could lead to a change in the position of the Fermi level relative to the conductive edge channels, which would then lead to a detectable change in electrical properties.    
Such a strong chemical interaction can be induced by starting from a defected GNR for adsorbing the molecule, having a single or double carbon vacancy. 
The preferential adsorption of related molecules, Cu-phthalocyanines, on graphene at defect sites has been demonstrated experimentally \cite{Dou2012}. 
Previous work addressed the stability of transition metal atoms on a double vacancy, and its role for conductance and spin filtering \cite{Tawfik2015a,Tawfik2016}. 
In this context, the present work may be perceived as a logical extension that addresses the issue how to place single transition metal atoms on the GNR: 
We demonstrate that exposing the GNR to metal porphyrins offers a route to decorate the GNR in a controlled way with single metal atoms. To achieve bonding, we employ as a starting geometry a GNR containing a single-atom vacancy.  Such a highly reactive defect can be generated by ion bombardment or by exposing the GNR to the high-energy electrons of a transmission electron microscope \cite{Vicarelli2015}.

In the present theoretical study, 
four transition metal porphyrins have been selected for an investigation of their adsorption properties. 
The choice was preceded by a screening of numerous metal porphyrins with respect to their chemical stability and their magnetic moment in the gas phase (see Supplementary Material). 
Both the physisorbed state on a perfect GNR, as well as the chemisorbed state at a GNR with a single-atom vacancy are studied by density-functional calculations including van der Waals interactions. Finally, the consequences of the adsorption for the electrical transport properties are studied using the non-equilibrium Green's function (NGEF) approach to conductance in nanoscale systems. The calculations enable us to address the change in magnetic moment upon adsorption, both at the transition metal centre and in the GNR. These changes also show up in the spin-dependent conductance. Each of the four porphyrins investigated, containing Ti, Nb, Fe or Ru, modify the spin polarization of the current in the nanoribbon in a characteristic way. The results of our calculations are of interest to graphene-based spintronics: Spin valves consisting of ferromagnetic metal electrodes on a GNR could be used to detect molecule-specific modifications of the spin-polarized current. 

\section{Methods}

The calculations presented in this work employ density functional theory (DFT) using the generalized gradient approximation (GGA-PBE) \cite{PeBu96} for the exchange-correlation functional.
This functional has been widely used in the description of extended systems such as graphene and its nanoribbons, as it is parameterized in such a way as to correctly recover the limit of an extended electron gas with weak spatial variations of the electron density. 
For localized electrons, in particular for the $d$ electrons of the transition metal atom in the porphyrin, hybrid functionals such as B3LYP could be more suitable. For this reason, some of the calculations have been repeated with the B3LYP functional.

The energetics of adsorption of metal porphyrins on a graphene nanoribbon are calculated using the 
all-electron code FHI-aims \cite{FHIaims}. This code works with numerical atom-centred basis functions
\footnote{The basis functions comprised the so-called 'light' basis set for C and H atoms, and the 'tight' basis set for the N atom and the transition metal, the latter two being involved in bond making and bond breaking.}, and can be applied to both non-periodic systems (molecules) as well as to periodic systems (nanoribbons). Moreover, it allows us to include 
van der Waals (vdW) interactions in an approximate yet efficient scheme using pair-wise, density-dependent interactions, as introduced in the work of Tkatchenko and Scheffler \cite{TkSc09}, in the following called TS vdW.  
This scheme has been demonstrated to give an excellent description of vdW interactions between a wide class of organic molecules \cite{Reilly2013}. 
Moreover, a self-consistent variant \cite{Ferri2015} of this scheme has become available recently that allows one to calculate, in addition to the total energy, the modifications of the charge density and the electronic band structure due to vdW interactions. 
Since the coefficients of the pair-wise interaction terms in this Tkatchenko-Scheffler scheme depend on the local electronic density, it adapts in a flexible way to many different density functionals that can be used to generate the ground state density. 
In addition to the PBE functional, the B3LYP hybrid functional \cite{Beck88,LeYa88,Beck93}  has been used, which introduces an admixture of the Fock energy in the description of exchange. Within the FHI-aims code, the calculation of exchange integrals is realized through auxiliary basis functions, following the resolution-of-identity approach \cite{FHIaimsHSE}.  

For the adsorption studies, we use a periodic geometry along the nanoribbon, with 10 rows of C atoms.
Thus, the formula for the perfect nanoribbon is C$_{70}$H$_{10}$. With these settings, we are able to model an adlayer of metal porphyrins at maximum coverage, i.e. with molecules nearly touching. The width of the GNR (7~C atoms) is sufficient for the molecule to interact over its full area with the GNR. 
For the calculations of electronic transport properties, it is required to add contacts to both sides of the GNR. 
In principle, these contacts should reflect the specific experimental conditions used to probe the electronic transport. 
Since the proposed experiment has not yet been performed up to now, we decide for the simplest choice -- contacts being formed by ideal, 7 C atoms wide GNRs of length of 4 atomic rows (red boxes in Fig.~\ref{fig:set-up}). 
In total, a larger periodicity of 18 rows of C atoms is used, i.e. the formula for the GNR supercell is now C$_{126}$H$_{18}$. 
The C and H atoms at the contact areas are held fixed at their ideal positions, while all atoms in the central region including the adsorbed porphyrin are allowed to relax. 
Previous calculations including vdW interactions for graphene \cite{Hamada2011} and for molecular adsorption systems \cite{Morbec2017} have shown that the dominant effect of the vdW interaction comes from changes of the geometry, while the changes of the electronic structure itself, once the correct geometry is employed, is of minor importance. 
Therefore, 
the C$_{126}$H$_{18}$+porphyrin system was structurally optimized with the GGA-PBE functional including vdW corrections to the forces on the atoms, and the relaxed geometry was passed to the transport calculation. 
For test purposes, also geometries optimized {\em excluding} vdW interactions were fed into the electronic transport code in order to monitor possible uncertainties arising from variations of the geometry. 
For the present system, it was found that the conductivities for both geometries (optimized with and without vdW interactions) showed qualitatively similar behavior (see Supplementary Material for details). Thus, the results are robust with respect to slight changes of the geometries.

Using the atomic structures obtained as described above, we perform 
non-equilibrium Green's function (NEGF) calculations as implemented in the TranSIESTA code~\cite{Brandbyge2002}. The single-particle Green's functions don't allow for an explicit treatment of electronic correlations; instead, these must be taken into account by the exchange-correlation potential which is part of the density functional. 
Since the vdW interaction affects the outcome of the calculations mostly via the atomic geometry and the electronic structure is nearly unaltered by the vdW correlations, we employ the conventional DFT-PBE functional within the TranSIESTA calculations. Moreover, single-$\zeta$ plus polarization basis sets were used for all of the atoms. 
The current $I_{\sigma}$ through the scattering region, for the spin channel $\sigma$, as a function of the bias voltage $V_{\textrm{\tiny{bias}}}$ across the device can be estimated using the Landauer-Buttiker formula~\cite{Datta1995},
\begin{eqnarray*}
\lefteqn{I_{\sigma}(V_{\textrm{\tiny{bias}}}) = } \\ & = & \frac{2e}{h}\int^{-\infty}_{\infty} \left[f_{L}(E-\mu_{L})-f_{R}(E-\mu_{R})\right]  T_{\sigma}(E,V_{\textrm{\tiny{bias}}})\, dE, 
	\label{LB}
\end{eqnarray*}
where $L$ and $R$ denote left and right electrodes, respectively, and $T_{\sigma}(E,V_{\textrm{\tiny{bias}}})$ is the transmission function for spin channel $\sigma$, which is a function of the energy ($E$) and $V_{\textrm{\tiny{bias}}}$, the voltage applied across the electrodes. $f_{L/R}$ is the Fermi-Dirac distribution function and $\mu_{L/R}$ is the electrochemical potential. 
The difference between the two Fermi-Dirac distribution functions in the left and the right lead occurring in Eq.~\ref{LB} indicates that an electronic state contributing to the current must be occupied in the left lead and unoccupied in the right lead, or {\it vice versa}. This is a necessary condition, the size of the contribution being determined by the transmission function 
$T_{\sigma}(E,V_{\textrm{\tiny{bias}}})$ given by the trace of the square of the transmission amplitude $\textbf{t}_{\sigma}$, which takes the form 
\begin{equation}
	T_{\sigma}(E,V_{\textrm{\tiny{bias}}})= \textrm{Tr}\left[\textbf{t}_{\sigma}^{\dagger}\textbf{t}_{\sigma}\right]=\textrm{Tr}\left[\Gamma_{R\sigma}G_{\sigma}\Gamma_{L\sigma}G_{\sigma}^{\dagger}\right] \, .
	\label{eq1}
\end{equation}
Here, $\Gamma_{R/L\sigma}$ is the imaginary part of the self-energy, and $G_{\sigma}$ is the Green's function of the scattering region.

\section{Results}

\begin{figure}[btp]
\begin{center}
\includegraphics[width=0.48\textwidth]{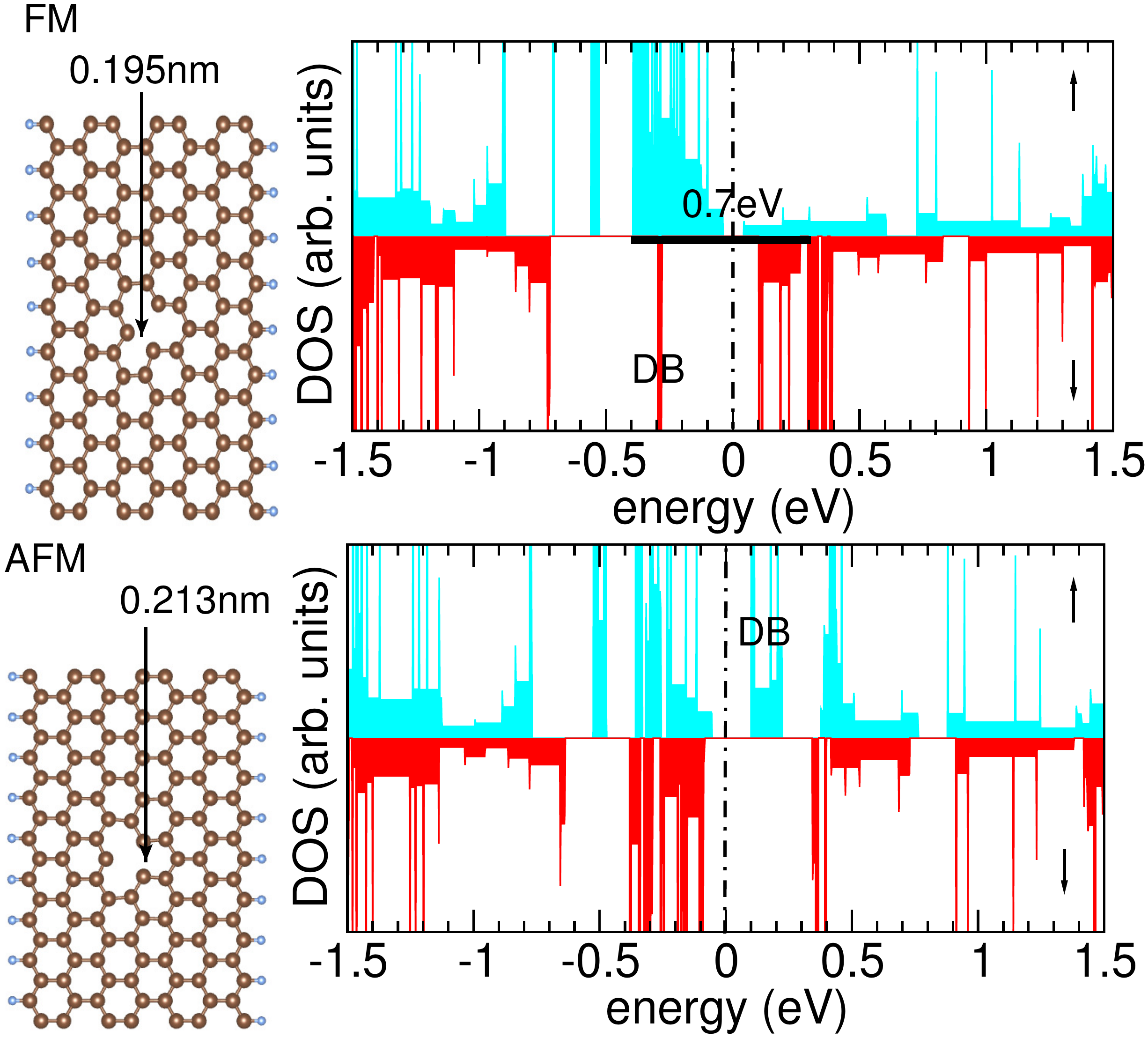}
\caption{Density of states obtained from the PBE + TS vdW functional for a zig-zag edge graphene nanoribbon with a single-atom vacancy for ferromagnetic spin alignment of the edge channels (upper part) and antiferromagnetic alignment (lower part). The Fermi energy is marked by the dashed vertical line at zero eV. The ball-and-stick models(left column)  display the different atomic relaxation in both structures. The FM structure is lower in energy by 0.22~eV. }
\label{fig:GNR}
\end{center}
\end{figure}
 
\subsection{Graphene nanoribbon with a single-atom vacancy}

Graphene nanoribbons with hydrogen-passivated zig-zag edges offer interesting transport properties, because they display spin-polarized edge channels.
In the present DFT calculations we find that the spin moment amounts to about $0.3 \mu_B$ at each edge C atom, plus smaller magnetic moments at C atoms in the interior of the GNR.  

In a defect-free GNR, the ground state is antiferromagnetic (AFM), i.e., there is a slight energetic preference (few meV per edge C atom) for antiparallel alignment of the spins at both edges, a result that is confirmed both in effective single-particle theories \cite{Son2006}  (such as DFT) as well as in a many-particle approach \cite{Jung2009}.
For the defected GNR, the energy difference between ferromagnetic (FM) and AFM arrangements of the edge magnetic moments is affected by symmetry considerations \cite{Li2008} that are accessible  at the DFT level of theory: 
Whereas the ideal GNR has either a mirror plane, for the case of an even width, or a glide plane, for the case of an odd width, introducing a single-atom vacancy in a GNR of odd width breaks this symmetry. 
As a consequence, we find that the electronic ground state is closely related to the relaxation of the atomic positions. 
The parallel alignment of the spins in the two edge channels is found to be energetically preferred, leading to a FM ground state. 
Concomitantly a relaxation of the atomic positions is observed that reduces the symmetry of the ideal GNR: two dangling bonds of C atoms next to the C vacancy form a weak bond oriented in an oblique direction (neither along nor perpendicular to the ribbon axis). These two atoms are only 0.195~nm apart (Fig. \ref{fig:GNR}, upper left), smaller than the ordinary distance between second-nearest-neighbor atoms of 0.247~nm. 
As we conclude from the one-dimensional band structure, the FM ground state results in an exchange splitting of about 0.7~eV between the edge channels of the spin-up and the spin-down electrons that is indicated in the density of states (DOS), see Fig. \ref{fig:GNR}, upper right. 
It results in only one edge spin channel to be occupied. 
The dangling bond (DB) of the remaining C atom pointing into the vacancy forms a localized state (marked by DB in Fig.~\ref{fig:GNR}) in the spin-down electrons. 
The asymmetry in the DOS can be understood by observing that the defected GNR lacks a symmetry plane along the ribbon direction. Hence, the magnetic moments inside the GNR, near the vacancy, are not fixed due to symmetry, but must be determined by a self-consistent calculation.

\begin{table}[ht]
\small
\caption{Adsorption energies (in eV) at full coverage, i.e. one molecule per 69 carbon atoms, obtained with the PBE + TS vdW functional. Left: physisorption on the perfect nanoribbon, Middle: chemisorption in `sandwich position' as shown in Fig.~\ref{fig:adsGeom}. The numbers in parenthesis are obtained from the B3LYP + TS vdW functional. 
Right: enthalpy $\Delta H$ (in eV) of the H-induced desorption reaction as discussed in the main text, leaving a transition metal atom at the vacancy in the GNR.}
\label{tab:Eads}
  \begin{tabular*}{0.48\textwidth}{@{\extracolsep{\fill}}lccr}
\hline
          & physisorption  & chemisorption & $\Delta H$(desorption) \\
\hline
Ti-porphyrin &  2.66 & 5.35  (5.18) & 1.17 \\
Nb-porphyrin &  2.58 & 7.27 (7.76) & $-0.04$ \\
Fe-porphyrin & 2.58 & 3.43  (2.93) & 4.00 \\
Ru-porphyrin & 2.69 & 4.59 (4.60) & 3.77 \\
\hline
\end{tabular*}
\end{table}

If two dangling bonds of C atoms near the vacancy form a bond pointing along the ribbon, an antiparallel alignment of the spins in the edge channel is the energetically preferred situation. For such a bond, we find a bond length of 0.213~nm (Fig. \ref{fig:GNR}, lower left). Because the upper and the lower end of the GNR are held fixed, there is less freedom for the atoms to relax if such a bond is formed. 
The AFM configuration for the seven-atom wide GNR C$_{125}$H$_{18}$ studied in this work is 0.22~eV, or 12~meV per edge C atom, higher in energy than the FM state. 
In the DOS of the AFM structure, both edge channels are occupied and very similar in their energetic position (Fig. \ref{fig:GNR}, lower right). 
The Fermi energy $E_F$ lies in a gap. Again the dangling bond of the remaining, unsaturated C atoms causes a non-dispersive state in the gap. However, in contrast to the FM configuration, this dangling-bond state (marked by DB in Fig.~\ref{fig:GNR}) is unoccupied in the AFM configuration. 
This is one of the reasons why the DOS of the spin-up and the spin-down electrons is not fully symmetric, as one would expect from bulk antiferromagnets, although the electron spins in the left and right edge channel are indeed antiparallel but equal in magnitude. 

\begin{figure}[htbp]
\begin{center}
\includegraphics[width=0.5\textwidth]{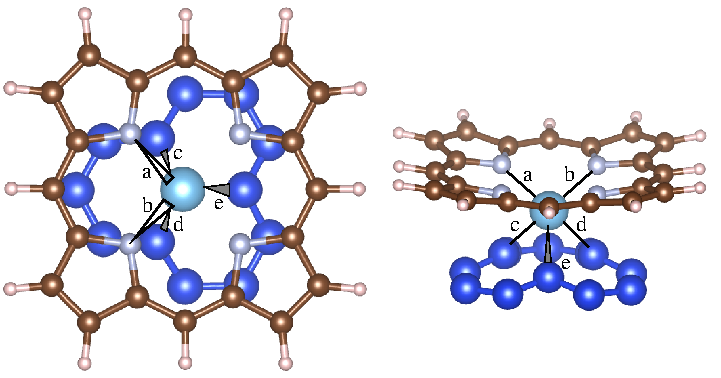}
\caption{Geometry of an adsorbed Ti-porphyrin on the single-atom vacancy of a zig-zag edge graphene nanoribbon. 
Only selected C atoms of the GNR are shown as blue (bigger) spheres. 
The Ti atom (light blue sphere) has two bonds (a,b) with two N atoms of the porphyrin ring, two short bonds (c,d) to the C atoms of graphene and one longer bond (e). 
The bond lengths indicated as a, b, c, d, e in the figure are given in Table~\ref{tab:geo} for all metal porphyrins studied.}
\label{fig:adsGeom}
\end{center}
\end{figure}

\subsection{Adsorption}

While we are not aware of experimental studies of porphyrin molecules on GNRs, 
the attractive interaction of a Co-porphyrin with graphene on a metal substrate has been observed in experiment \cite{Klar2014}. 
Moreover, similar organic ring molecules have been adsorbed on a graphene sheet grown epitaxially on metal substrates. 
One example is the experimental study of 
Cu-phthalocyanine on graphene \cite{Xiao2013}.
For this system \cite{Cardenas-Jiron2011}, and likewise for Cu-phthalocyanine \cite{RenKaxiras2011} and tetraphenyl-porphyrin \cite{Malic2014} on graphene, 
theoretical studies using DFT plus pairwise van der Waals interactions have been performed.  
These works find that the molecules are bound in the physisorbed state, i.e. a chemical interaction between graphene and these molecules is absent.

Analogously, we find that, on an ideal graphene nanoribbon, physisorption is also the prevailing interaction for the porphyrin molecules investigated in the present paper.
The adsorption height of 3.4~{\AA} is found to be very similar for all four molecules. 
This is already a strong indication that the attractive interaction is mostly due to the van der Waals interaction (here described by the Tkachenko-Scheffler pairwise interaction functional \cite{TkSc09}), while 
the PBE-GGA functional \cite{PeBu96} makes a slight repulsive contribution (of the order of 0.2 eV to 0.4 eV) at the adsorption height.  
The binding energies  of physisorption are given in Table~\ref{tab:Eads}. 
They are calculated from the energy differences between the combined system and the single-atom vacancy GNR plus the metal porphyrin taken separately. In these calculations, all three structures are relaxed, and for the metal porphyrin in the gas phase, the ground-state multiplet, as specified in the Supplementary Material, is used. 
As already observed for the adsorption height, the physisorption energies are all very similar. It is found that the heavier element Ru yields a slightly stronger binding due to its higher polarizability than the lighter Fe. For Ti, the adsorption energy is slightly larger than for Nb, which could indicate that the bonding is not purely of van der Waals type, but charge transfer and ionic bonding start to play a role in the case of Ti.

\begin{table}[bhtp]
\small
\caption{Adsorption height $h$ with respect to the plane of the graphene nanoribbon, and bond length (in \AA ) between the metal atom and N atoms (a,b) and C atoms (c,d,e) of metal porphyrins chemisorbed on a single-atom vacancy. 
The early transition metals are four-fold coordinated (the third C neighbour is further away), while the late transition metals Fe and Ru are five-fold coordinated (c = d = e). The quoted results are obtained with PBE + TS vdW interaction.}
\begin{tabular*}{0.48\textwidth}{@{\extracolsep{\fill}}lcccc}
\hline
  & $h$ & a,b  & c,d  & e \\
   \hline
   Ti-porphyrin & 2.33 &  2.22  & 2.15 & 2.28 \\
   Nb-porphyrin  & 2.24 & 2.26 & 2.14  & 2.31 \\
      Fe-porphyrin & 1.88 &  2.15  & 1.92  & 1.92 \\
   Ru-porphyrin & 2.01 &  2.24  & 1.99  & 1.99  \\ 
\hline
\end{tabular*}
\label{tab:geo}
\end{table}

Our next goal is to study {\em chemical} interactions between porphyrins and a GNR beyond mere physisorption. 
Placing the molecule above a GNR with a single C vacancy,   
we are able to induce an exothermic reaction leading to chemisorption. 
As shown in Table~\ref{tab:Eads}, the reaction enthalpy is strongly element-specific, spanning a range of 3.43~eV for Fe to 7.27~eV for Nb.
We obtain a sandwich geometry, where the metal atom sits at intermediate height between the nanoribbon plane and the molecular plane (cf. Fig.~\ref{fig:adsGeom}). 
While in the free molecule all four metal--N bonds are of equal length, we notice that, upon chemisorption, two adjacent metal-N bonds are broken, leaving the two metal-N bonds (labelled a and b in Fig.~\ref{fig:adsGeom} and Table~\ref{tab:geo}) intact.
At the same time, the molecule is no longer planar. The detached N atoms move further away from the nanoribbon.
These findings indicate that the planar, aromatic metal porphyrin molecules are quite stable, and the aromaticity must be broken before chemisorption becomes possible.
Therefore one would expect that chemisorption requires the molecule to overcome some activation energy barrier. 
However, since we could locate the chemisorbed state simply by placing the molecule above the vacancy, such a barrier should not be very high.
While the early transition metals Ti and Nb form shorter bonds (c and d) to two C atoms of the GNR and one slightly longer bond, Fe and Ru form equal bonds to three C atoms (c,d, and e). The five-fold coordination of Fe and Ru in the sandwich structures formed at the defected GNR can be understood on the basis of coordination chemistry in gas-phase metal complexes, and corresponds to the well-known five-fold-coordinated complexes Fe-pentacarbonyl and Ru-pentacarbonyl. In coordination chemistry, the particular stability of these complexes is usually explained by a full shell of 18 valence electrons, eight of which stem from the $s$ and $d$ shells of the metal atom, while ten more electrons are contributed by the lone pairs of the ligands. In the present case, their role is taken by two N and three C ligands. 
In the early transition metal atoms, such as Ti and Nb, their energetically higher-lying $3d$ or $4d$ electrons make shell-filling arguments inapplicable.

\begin{table}[htb]
\small
\caption{Magnetic moment (in $\mu_B$) for the whole structure (tot) of nanoribbon (18 C rows in length) plus molecule, and at the transition metal atom ($M$). $\Delta E$ gives the energy difference between the antiferromagnetic (AFM) state and the ferromagnetic (FM) ground state. Calculations are done for the chemisorbed sandwich structure (see Fig.~\ref{fig:adsGeom})  using PBE + TS vdW interaction.}
\begin{tabular*}{0.48\textwidth}{@{\extracolsep{\fill}}lccccc}
   \hline
   &   \multicolumn{2}{c}{FM}  &   \multicolumn{2}{c}{AFM}  & $\Delta E$ \\
   &  $M_{\rm tot}$  &  $M$  &   $M_{\rm tot}$  &  $M$  & (eV) \\
   \hline
7C-GNR+vacancy & 6.00  & -- &  0.00 & -- & 0.22 \\
+Ti-porphyrin  &   4.25 & 0.00  & 0.25    &   0.00 & 0.22  \\   
+Nb-porphyrin  &  4.82 & 0.02  & 0.92    & 0.04  &  0.00  \\ 
+Fe-porphyrin  &  6.30   &  1.58   &  1.80  & 1.47   & 0.09 \\ 
+Ru-porphyrin  &  5.95  &   0.86  &  1.29   &  0.78  & 0.02 \\  
   \hline
\end{tabular*}
\label{tab:magMom}
\end{table}%
In order to check the validity of the PBE functional for the molecules of interest that contain localized $3d$ or $4d$ electrons, the chemisorbed state was calculated (including atomic relaxation) using the B3LYP functional augmented by the TS treatment of vdW interactions. Hybrid functionals such as B3LYP are generally expected to give a better description of localized electrons due to their admixture  \cite{Beck88} of exact exchange. 
As Table~\ref{tab:Eads} shows, the adsorption energies obtained with B3LYP are in the same range (within 0.5 eV) as the PBE results. This gives us confidence that the valence state of the transition metal is described correctly in our DFT calculations.

\begin{figure*}[tbp]
\begin{center}
\includegraphics[width=0.7\textwidth]{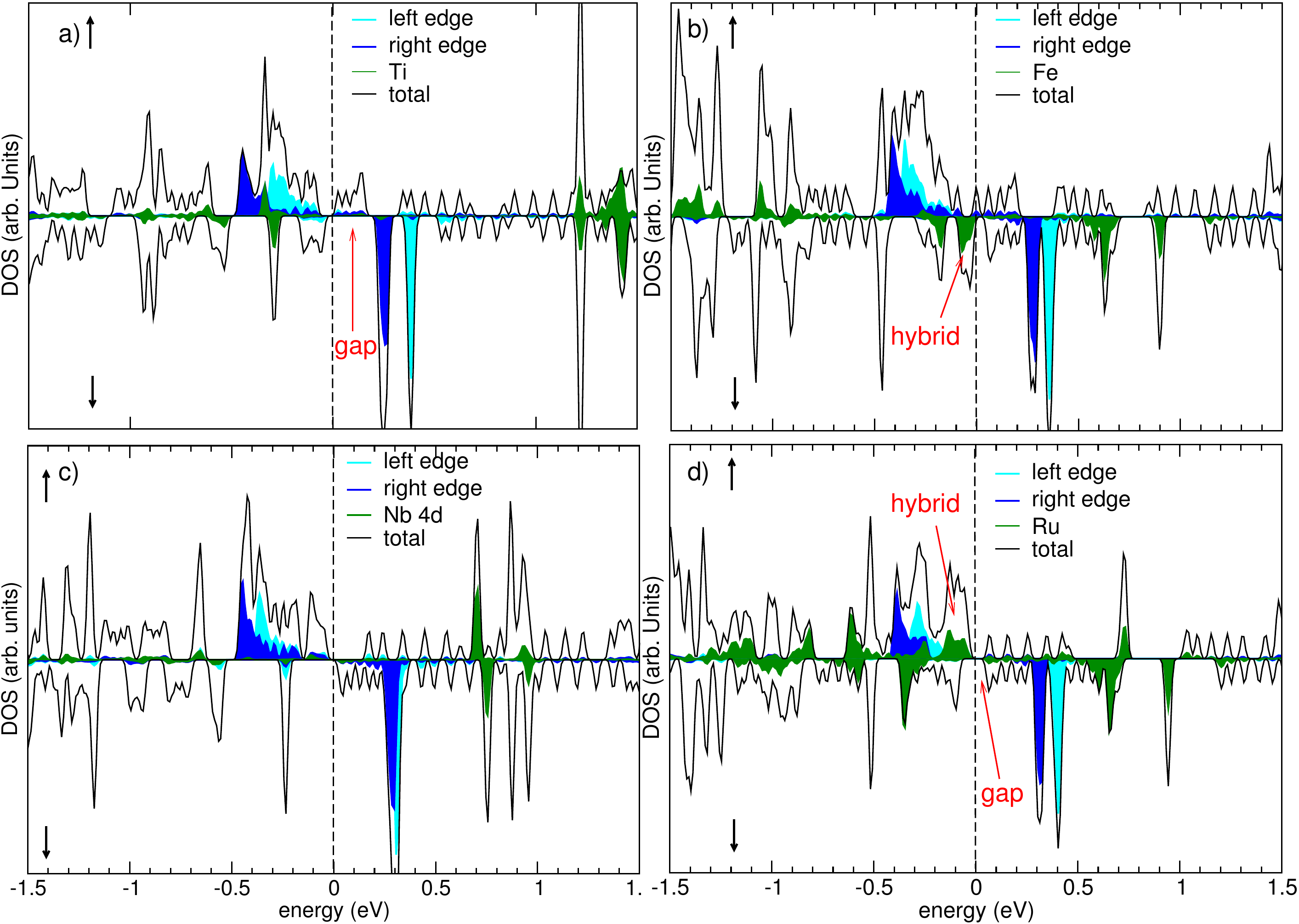}
\caption{Density of states obtained from the PBE + TS vdW functional for the combined system of a metal porphyrin adsorbed on a single-atom vacancy in a zig-zag edge graphene nanoribbon. 
The spin-up DOS corresponds to positive values, the spin-down DOS to negative values. 
The Fermi energy is indicated by the dashed vertical line at zero energy. 
The upper row shows the DOS for Ti-porphyrin (left) and Fe-porphyrin (right).
The lower row shows the DOS for Nb-porphyrin (left) and Ru-porphyrin (right). 
}
\label{fig:PDOS}
\end{center}
\end{figure*}

Moreover, test calculations were performed with the goal to remove the organic ring structure from the nanoribbon after chemisorption of the metal atom. 
However, the metal-free product formed in a possible desorption, the C$_{20}$H$_{12}$N$_4$ molecule, is highly unstable. 
Therefore, the envisaged desorption would be very costly in terms of energy. 
It is energetically less costly to desorb the whole molecule {\em including} its transition
metal centre, see Ref.~\cite{Xiao2013}. 
The picture changes if we consider a reaction of the adsorbate with hydrogen atoms. Then, the porphine molecule C$_{20}$H$_{12}$N$_2$(NH)$_2$  (also called the 'free base')  can form, which is about 10~eV more stable than the empty ring. 
The reaction enthalpy $\Delta H$ for the desorption reaction (M-porphyrin)$_{\rm ads}$ + 2~H $\to$ porphine  + M$_{\rm ads}$ + $\Delta H$ is given in the rightmost column in Table~\ref{tab:Eads}. The net reaction is almost thermoneutral for the Nb porphyrin, and strongly exothermic in all other cases. 
Therefore, exposure to hydrogen could be one possibility to chemically remove the organic ring structure from the GNR if desired, leaving behind the metal atom in the vacancy. 
Thus, exposing defective graphene or graphene nanoribbons to metal porphyrins, followed by a hydrogen treatment, allows one to deposit a single metal atom in a controlled way avoiding metal agglomeration.

By comparing the magnetic moments of the transition metal atoms in the chemisorbed state (Table~\ref{tab:magMom}) with those in the free molecule (see Supplementary Material) 
we notice that chemisorption goes along with a loss or a reduction of magnetization. 
Ti and Nb loose their magnetic moment completely, while  it is partly retained by Fe and Ru. 
The $3d$ metal Fe  preserves more of its magnetic moment compared to the $4d$ analog Ru.
Remarkably, the FM alignment of the magnetic moments on the GNR remains to be energetically favourable also after chemisorption of the molecules. 
For the Ti-porphyrin, the preference for the FM state is strongest, while it is reduced for the Fe- and Ru-porphyrins. Only for Nb-porphyrin adsorption, the AFM and FM alignment are energetically degenerate within the accuracy of our calculations. 
Concerning the absolute magnitude of the magnetic moment at the Fe and Ru atoms, a slightly larger (by about $0.1 \mu_B$) moment is retained in the FM than in the AFM state. 
The mutual re-enforcement of the magnetic moments in the GNR edge states and at the metal centre may explain why the FM state is still preferred after the adsorption of Fe- and Ru-porphyrin.

\subsection{Electronic properties}

Insight into the chemical bonding and the resulting electronic and magnetic properties of the chemisorbed state can be gained from inspecting the orbital-projected density of states shown in Fig.~\ref{fig:PDOS}. 
It is found that the overall electronic structure of the single-vacancy GNR (cf. Fig.~\ref{fig:GNR}) is still largely retained even after chemisorption. 
The edge channels are still present, separated by an exchange splitting of 0.7 eV into an occupied majority-spin (= up spin) and an unoccupied minority-spin (= down spin) channel. 
However, the exact position of the Fermi energy between the exchange-split edge states varies among the different adsorbates. 

In the chemisorbed state, the early transition metals Ti and Nb have given away their $3d$ and $4d$ electrons, respectively, to bonding orbitals. 
This can be seen from the unoccupied $3d$ or $4d$ states (green curves)  in the left column of Fig.~\ref{fig:PDOS}, and is consistent with the loss of their magnetic moment reported in Table~\ref{tab:magMom}. 
We find that Ti-porphyrin adsorption has only a minor effect on the electronic structure of the GNR around the Fermi energy: 
The gap in the minority spin observed in the FM ground state (cf. Fig.~\ref{fig:GNR}, upper right) of the single-atom vacancy GNR persists. 
For Nb, having one more valence electron than Ti, the Fermi level  in the minority spin lies directly above this gap, allowing for one more electron to be accommodated.
The adsorption of Fe- and Ru-porphyrin has a more profound effect on the electronic structure. 
The free molecules have unoccupied $3d$ or $4d$ states in the minority spin. Upon adsorption, these states hybridize with electronic states of the GNR, and form both occupied bonding states and unoccupied states.  For Fe, occupied states of Fe $3d$ character are observed at 0.1~eV and 0.2~eV below $E_F$, and unoccupied states are found at 0.6~eV and 0.9~eV above $E_F$ (Fig.~\ref{fig:PDOS}, upper right panel, green curve). 
The minority-spin $4d$ states of Ru are split into bonding states that are located 0.4~eV below $E_F$, and 
unoccupied minority spin states located 0.6~eV and 0.9~eV above $E_F$ (Fig.~\ref{fig:PDOS}, lower right panel, green curve). 
Consequently, for both Fe and Ru, electronic states appear close to $E_F$ in both spin channels. This is qualitatively different from the situation before adsorption, where a clear gap was present in the minority spin of the single-vacancy GNR.

\begin{figure}[tbp]
\begin{center}
\includegraphics[width=0.5\textwidth]{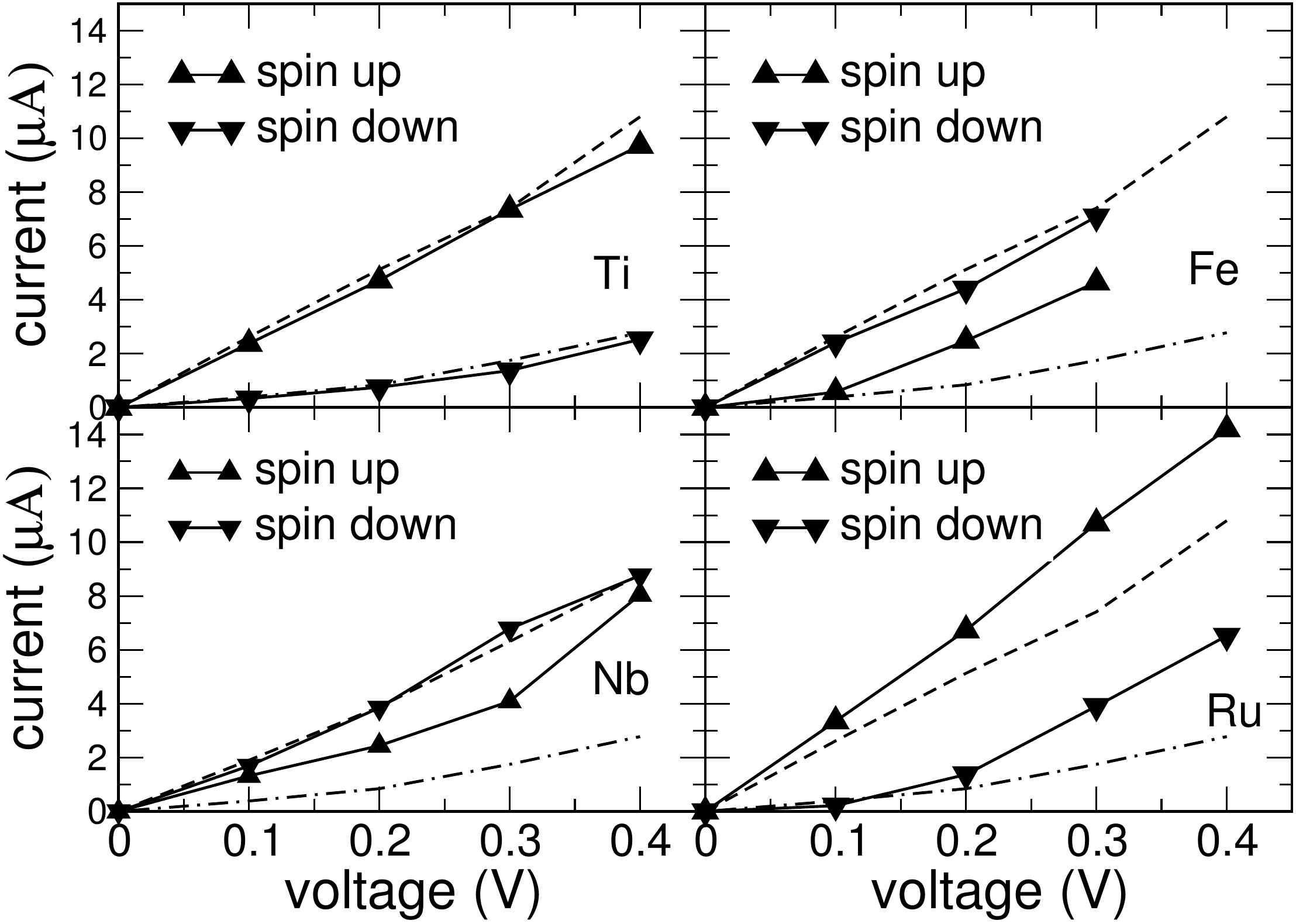} 
\caption{Spin-resolved current (in $\mu$A) as function of source-drain voltage for the combined system of a metal porphyrin adsorbed on a single-atom vacancy in a zig-zag edge graphene nanoribbon. 
The upper row shows the currents for Ti-porphyrin (left) and Fe-porphyrin (right), the lower row for Nb-porphyrin (left) and Ru-porphyrin (right).
The lines with triangular symbols show the currents in the spin channels. 
For comparison, the spin currents of the clean GNR with a single-atom vacancy are shown as dashed (spin-up) and dash-dotted (spin-down)  lines in all subfigures. 
}
\label{fig:transport}
\end{center}
\end{figure}

\subsection{Transport properties} \label{sec:transport}

The NEGF calculations have been performed with the TranSIESTA code using  the relaxed atomic geometries from the previous section as input. 
In our standard procedure, the vdW interactions have been taken into account when determining the adsorption geometry.  
The current-voltage (I-V) characteristics for these geometries are presented in Fig.~\ref{fig:transport}.  

For the GNRs with a metal porphyrin adsorbed, we additionally performed calculations for the adsorption geometry obtained with the PBE functional only, {\em disregarding} the vdW interactions, in order to assess the sensitivity of the results to geometrical details. 
Despite clear differences in the adsorption height of the molecule, 
we noticed that the qualitative behaviour of the I-V characteristics is very similar for both geometries 
(see Supplementary Material). 
This can be understood as an indication that the electric current is carried mostly by electronic states near the edges of the GNR, thus avoiding the adsorption site. 
However, each molecule has a specific effect on the energetic positioning of the Fermi level in the scattering region. Therefore, we observe clear differences in the conductive behaviour between different molecules, while the inclusion of the vdW interaction between molecule and GNR has only a modest effect on the conductance within the GNR.

For the clean single-vacancy GNR (dashed lines in Fig.~\ref{fig:transport}), we find that the current is strongly spin-polarized and dominated by the majority-spin channel. 
This can be rationalized by the Fermi energy falling into a gap in the minority-spin density of states (Fig.~\ref{fig:GNR} upper right). 
For the GNR with adsorbed porphyrin,  considerable spin polarization of the current is observed, except for Nb-porphyrin.
In the dominant spin channel, it can be seen from the curves in Fig.~\ref{fig:transport} that current and voltage are nearly proportional to each other;
i. e., the transport is Ohmic. 
In the minority spin channel,  activated behaviour of the conductance, i. e. a super-linear rise of the current with voltage, is observed for Ti-, Fe- and Ru-porphyrin. For both adsorbates, the minority channel contributes very little at 0.1~V, more than an order of magnitude less than at 0.4~V, the highest voltage studied. 
This activated behaviour is found to be correlated with a gap near $E_F$ in the DOS of the minority spin channel, 
see Fig.~\ref{fig:PDOS}a) and d).

\begin{figure}[tbhp]
\begin{center}
\includegraphics[width=0.5\textwidth]{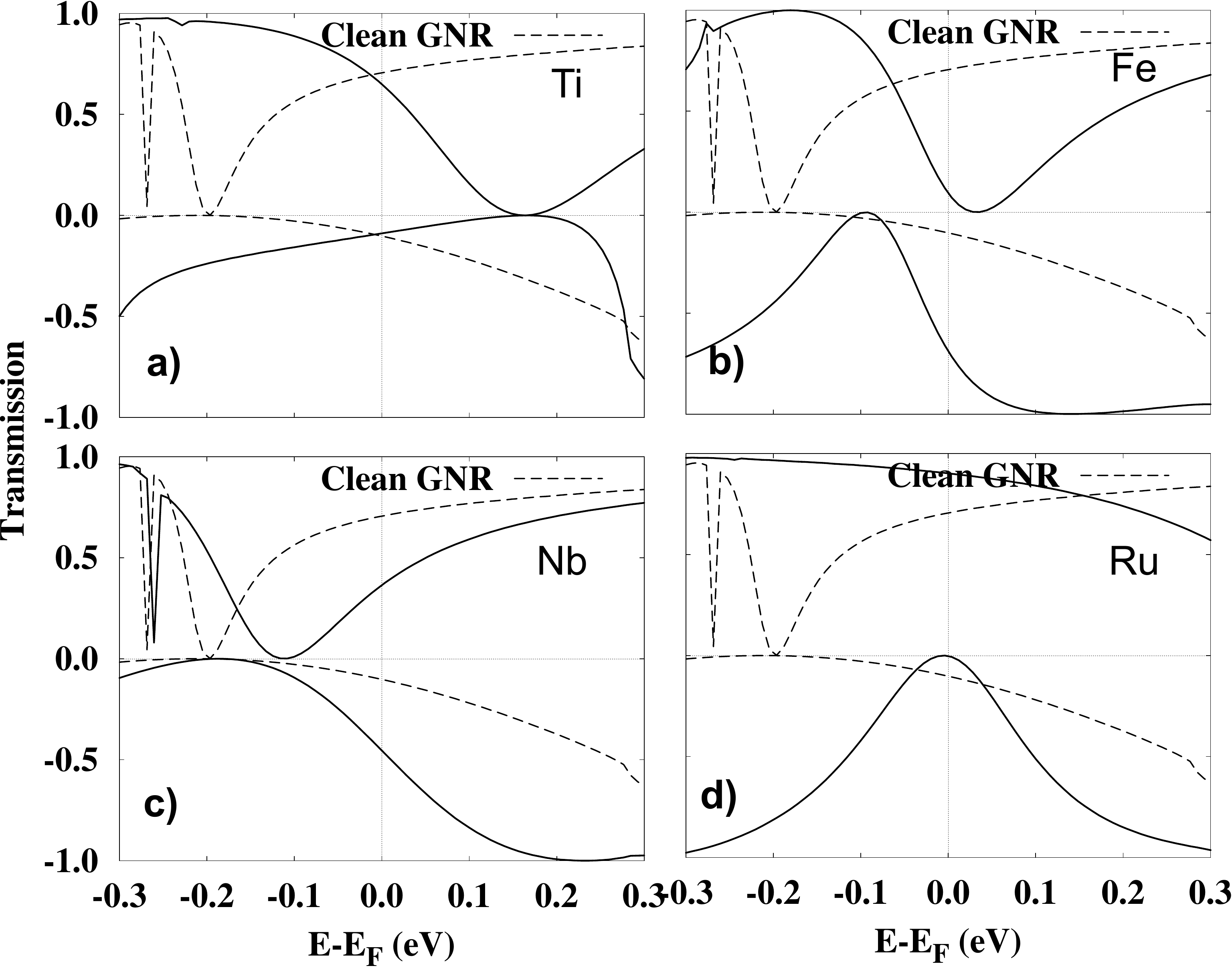} 
\caption{Spin-resolved transmission function, $T_{\alpha}(E,V_{\rm bias})$ at $V_{\rm bias} = 0.1$~V for the combined system of a metal porphyrin adsorbed on a single-atom vacancy in a zig-zag edge graphene nanoribbon. 
The full lines show the transmission function for the (a) Ti, (b) Fe, (c) Nb and (d) Ru. The dashed-lines show the transmission function for the clean GNR.}
\label{fig:transmission}
\end{center}
\end{figure}

Further insight into the origin of the I-V characteristics can be gained by investigating the spin-resolved transmission functions, $T_{\alpha}(E,V_{\rm bias})$ for the considered structures, Fig.~\ref{fig:transmission}. We focus here on the $T(E,V_{\rm bias})$ at $V_{\rm bias}=0.1 $ V, because this is where the absolute magnitude of the spin-filtering is highest in Ti-, Fe-,  Ru-porphyrin, as well as for the clean GNR. The bias window (that is, the energy window in which the transmission is integrated in order to obtain the current $I$, as in Eq. ~\ref{LB}) goes from $-V_{\rm bias}/2$ to $+V_{\rm bias}/2$. We display the transmission of the adsorption systems in comparison with the clean GNR in each plot. 
For the clean GNR, there exists a broad range of energies with very low transmission  in the spin-down channel, whose minimum is at about $-0.2$ eV, and thus further below the Fermi energy than in any of the adsorption systems. Even so, the suppression in the transmission in the spin-down channel is sufficiently strong to lead to significant spin filtering of the clean GNR.

Next, we discuss the various cases of adsorbed porphyrin molecules. 
It is observed that the I-V characteristics for Ti- and Ru-porphyrin show similar trends as for the clean GNR. 
This indicates that, similar to the clean GNR, the edge channels are still providing the major contribution to transport in these two systems.
In the  transmission (Fig.~\ref{fig:transmission}a) and d) ), there is a dip close to the Fermi energy in the spin-down transmission channel 
resulting in a net current that is polarized in the spin-up direction. 
For Ru, the conductance in the majority channel is even above the clean GNR, while it is slightly below it for Ti.
The former observation can be understood by inspecting Fig.~\ref{fig:PDOS}d) showing  newly introduced states near $E_F$ formed via hybridization of the Ru $4d$ orbitals with the electronic states of the GNR. These make an additional contribution to majority-spin conduction.  

Adsorbed Fe-porphyrin behaves qualitatively different, since it displays a transmission dip right at the Fermi energy in the spin-up channel, which can be attributed to destructive quantum interference. 
Under this condition, the newly introduced hybrid states with Fe $3d$ orbital character in the minority spin channel (see Fig.~\ref{fig:PDOS}d)) dominate the transport. This is why the spin-filtering in the Fe-system is in the opposite direction to the other systems.  

For Nb-porphyrin, both spins contribute almost equally to the conductance. 
Here, the Fermi energy $E_F$ has been shifted so much by the adsorption that states in the majority / minority spin channel below / above $E_F$ contribute both to transport. 
This is corroborated by Fig.~\ref{fig:transmission} (c) where it is seen that Nb-porphyrin adsorption balances the transmission through each spin channel and therefore there is minimal spin polarization in the current. 

\begin{figure}[btp]
\begin{center}
\includegraphics[width=0.4\textwidth]{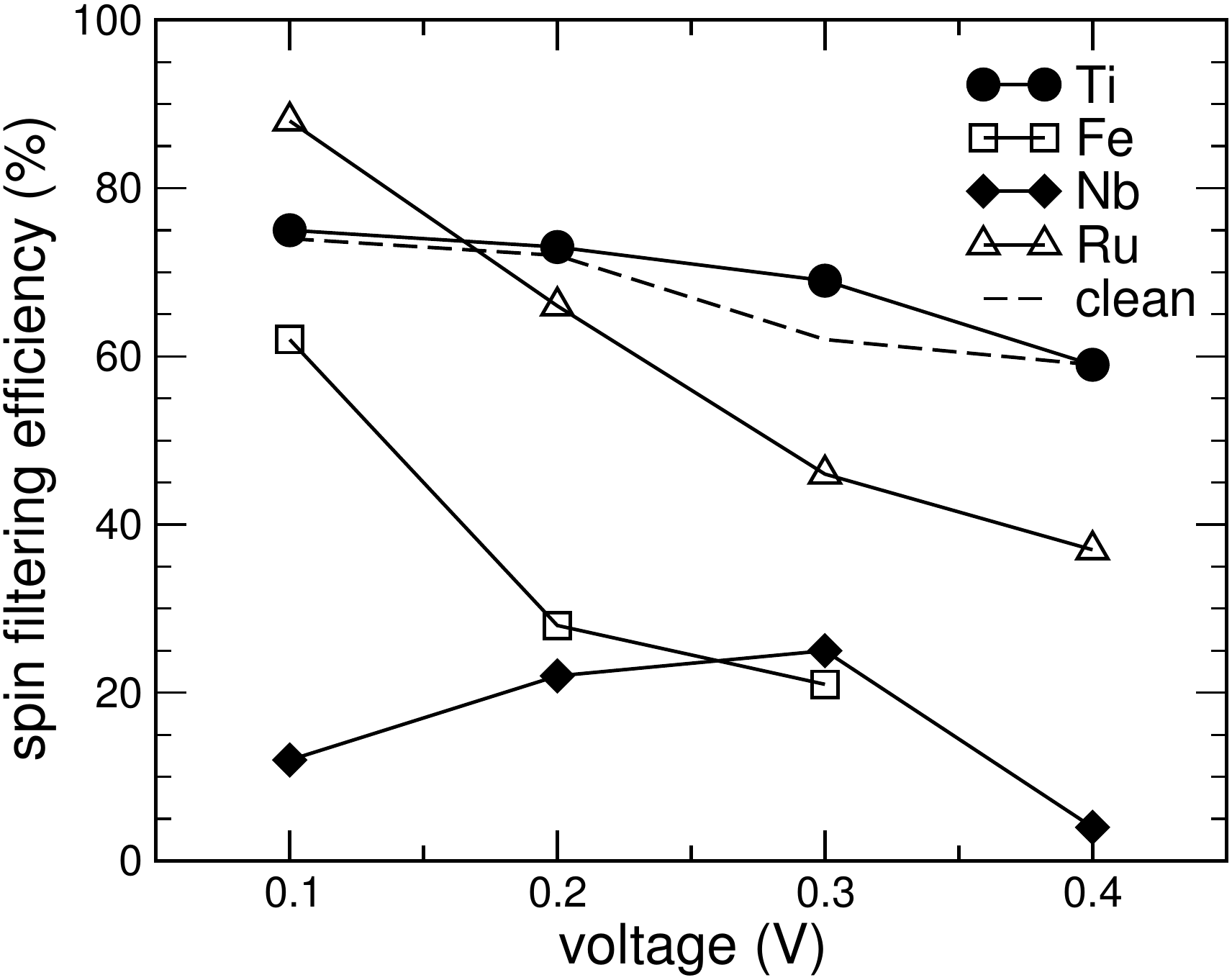} 
\caption{Spin filtering efficiency as function of the source-drain voltage for a graphene nanoribbon with a single-atom vacancy (dashed line) and with   
various metal porphyrins adsorbed on it (symbols).}
\label{fig:spinpol}
\end{center}
\end{figure}

For spintronics applications, e.g. in spin filters, obtaining a low conductance in the minority spin channel is of particular importance.  
As can be seen from Fig.~\ref{fig:transport}, however, by adsorbing Fe-, Nb- or Ru-porphyrin, its value  is more than doubled compared to the clean GNR.  
Only Ti is able to bring the conductance in the minority spin channel slightly below the value of the pure, single-atom vacancy GNR. 
In order to obtain a figure of merit, we define the spin filtering efficiency by $| I_{\uparrow} - I_{\downarrow}|/| I_{\uparrow} + I_{\downarrow} |$,
where $I_{\uparrow}$ and $I_{\downarrow}$ refer to the current in the spin-up and the spin-down channels, respectively. It is displayed as a percentage in Fig.~\ref{fig:spinpol}. 
The effect of spin filtering is already quite high (in the range of 60\% to 70\%) in the clean GNR with the vacancy. Adsorption of Ti-porphyrin has a very small effect on the states near $E_F$ and thus preserves the spin-filtering property. The opposite extreme is Nb-porphyrin, which upon adsorption creates conductance in both spin channels, and thus strongly reduces the spin filtering. Thus, the difference between Nb and Ti of just one valence electron obviously has a drastic effect: the increased valence electron count forces $E_F$ to move out of the gap in the minority-spin DOS (see Fig.~\ref{fig:PDOS}c) ). Moreover, since the FM and AFM states are energetically close for the case of Nb, fluctuations of the magnetization could further wash out the differences between the spin channels. 
Since Fe and Ru introduce states near $E_F$ in the minority spin channel, both Fe- and Ru-porphyrin tend to decrease the spin filtering efficiency at high voltages of 0.3 to 0.4~V. At 0.1~V, however, the spin-filtering in a GNR with Ru adsorbed can even be more efficient than for the pure GNR. This is because the conductance of the minority spin states shows activated behaviour; the states introduced directly at $E_F$ stem from the porphyrin molecule and thus are mostly localized states, contributing little to the conductance. 
For Ru-porphyrin at a voltage of 0.1~V, this effect leads to a spin filtering efficiency of $88$\%, even higher than for Ti-porphyrin.

Finally, we note that spin was treated as a conserved quantity in our transport calculations, i.e. the possibility of spin-flip scattering of electrons \cite{Wilhelm2015} from the Fe and Ru magnetic moments was not included. Studying the effects of inelastic scattering on transport could be an interesting topic for a future extension of this work, but is outside the scope of the methods used in this study.

\section{Conclusion}
We have performed first-principles calculations of the adsorption and detection of metal porphyrins on a single-atom vacancy in a graphene nanoribbon. 
The results show that the chemical interaction of the porphyrins with a single-atom vacancy is strong enough to break the aromaticity of the molecule and to induce chemical bonding. 
The chemisorption leads to subtle, but element-specific changes of the electronic structure of the nanoribbon that can be detected in transport measurements. 
Since the spin channels in the nanoribbon are affected differently by the adsorption, detecting not only the magnitude, but also the degree of spin polarization of the current  gives access to additional information that can be used to distinguish between adsorbed porphyrins with different metal centres. For instance, we find that adsorption of Ti-porphyrin preserves the high spin filtering capability of the nanoribbon, while Nb-porphyrin destroys it. 
This proposed method of element-specific detection is sensitive enough to discriminate even between metal centre atoms from the same group, but different row of the periodic table, such as Fe and Ru. 
Due to the extremely long spin diffusion length in graphene and the possibility to use ferromagnetic contacts, it appears plausible that a spin valve made of two contacts of a ferromagnetic metal (switchable from parallel to antiparallel magnetization) can indeed be used to read out this information encoded in the spin degree of transport. In this set-up, a varying  external magnetic field could be used to switch the magnetic orientation of the contacts. Since the exchange interaction between the graphene nanoribbon and the magnetic moment of the metal centre in the porphyrin is in most cases (Ti, Fe, Ru)  quite robust, the external magnetic field will still be too weak to affect the ferromagnetic alignment of the spins between the metal centre of the porphyrin and the  graphene nanoribbon. Thus, the mechanism governing the spin polarization of the current will not be disturbed by the detection apparatus. Moreover, our calculations point to an elegant method of depositing single metal atoms on a vacancy in graphene by offering a metal porphyrin followed by a hydrogen treatment. After chemical reaction and desorption of porphine, the transition metal centre of the porphyrin will be left behind, filling the vacancy with a metal atom. 

\section*{Acknowledgements}

This research was undertaken with the assistance of resources from the National Computational Infrastructure (NCI), which is supported by the Australian Government.

\section*{Supplementary Material}
\subsection{Stability of metal porphyrins in the gas phase}

Both the porphine, as well as the metal porphyrins in their ground states, are aromatic systems that show considerable stability.
Since we are interested in anchoring the porphyrin on a single-atom vacancy in graphene, and thereby the metal centre needs to move out of the porphine ring, knowing the binding strength of the metal centre in gas-phase porphyrins is a essential. Therefore, 
as a first step, the stability of metal porphyrins in the gas phase is studied using the code DMol$^3$ \cite{Dell00} that employs atom-centred numerical basis functions for the orbitals. 

The results of the calculations for free porphyrin molecules are summarized in Table~\ref{tab:survey}. 
The exchange reaction of a free metal atom with porphine, releasing two H atoms, is endothermic in the case of the noble metals Cu and Ag, and the simple divalent metals Mg, Ca and Zn. 
In the transition metal series, the binding of  the two H atoms is weaker than for a metal atom from the very left of the transition metal series (e.g. for Ti and V) and from the very right (Fe, Co and Ni), whereas Mn, being located in the middle of the series, binds more weakly than the two H atoms, such that the  exchange reaction becomes endothermic. 

\begin{table}[htp]
\small
\begin{center}
  \begin{tabular*}{0.48\textwidth}{@{\extracolsep{\fill}}lrrrrr}
\hline
   & $M_{\rm mol}$   & $M_{\rm at}$ & $M_{\rm at}^{\rm free}$ & $\Delta E$ &  $\Delta z$ \\ 
\hline
Ag & 1 & 0.42 &1 & 4.11 & 0 \\
Ca & 0  & 0 & 0 & 0.45 & 1.20 \\
Co & 1 & 1.08 & 3 & $-1.53$ & 0 \\
Cr & 4  & 3.96 & 6 & $-0.31$ & 0 \\
Cu & 1  & 0.57 & 1 & 1.75 & 0 \\
Fe &  2 & 2.11 & 4 & $-0.76$ & 0 \\
Mg &  0 & 0 & 0 & 0.84 & 0 \\
Mn & 3 & 2.98 &5 & 0.15 & 0 \\
Mo & 4 & 3.65 & 6 & $-0.15$ & 0 \\
Nb & 3  & 2.24 & 5 & $-1.18$ & 0.68 \\
Ni & 0  & 0 & 2 & $-1.24$ & 0 \\
Pd & 0  & 0 & 0  & $ 0.87 $ &0 \\
Rh & 1 & 1.04 & 3 & $-0.66$ & 0 \\
Ru & 2 &1.90 &4 & $-0.70$ &0 \\
Sc & 1 & 0.18 & 1 & $-1.68$ &0.55 \\
Tc & 3 & 2.71 &5 & $-0.87$ & 0 \\
Ti & 2 & 1.45 & 2 & $-1.78$ & 0 \\
V & 3 & 2.79&  3 &  $-1.89$ & 0 \\
Y  & 1 & 0.07 & 1 & $-1.46$ & 1.04 \\
Zn & 0 & 0 & 0 & 3.18  & 0  \\
Zr & 2 & 1.27 & 2 & $-2.00$ & 0.80 \\
\hline
\end{tabular*}
\caption{Basic physicochemical quantities of metal porphyrin molecules. 
The first column gives the elemental symbol, the second and third columns the magnetic moments in $\mu_B$ for the whole porphyrin molecule and at the metal centre, respectively. The fourth column gives the  magnetic moments in $\mu_B$ for the free metal atom according to the Hund's rule. The fifth column gives the binding energy $\Delta E$ of the metal centre, i.e. the energy (in eV) of the reaction M-porphyrin + 2 H $ \to$ porphine + M. The rightmost column gives the distance in {\AA} of the metal centre from the porphyrin plane if the out-of-plane configuration is the more stable one. The results were obtained with the PBE functional using the DMol$^3$ code\cite{Dell00}. }
\label{tab:survey}
\end{center}
\end{table}

 \begin{table}[tbp]
\small
\begin{center}
\begin{tabular*}{0.48\textwidth}{@{\extracolsep{\fill}}lcccc}
\hline
  & $h$ & a,b  & c,d  & e \\
   \hline
   Ti-porphyrin & 2.64 &  2.23  & 2.16 & 2.28 \\
   Nb-porphyrin  & 2.56 & 2.28 & 2.15  & 2.32 \\
      Fe-porphyrin & 2.05 &  2.14  & 1.92  & 1.91 \\
   Ru-porphyrin & 2.23 &  2.24  & 2.00  & 1.99  \\ 
\hline
\end{tabular*}
\caption{Adsorption height $h$ with respect to the plane of the graphene nanoribbon, and bond length (in \AA ) between the metal atom and N atoms (a,b) and C atoms (c,d,e) of metal porphyrins adsorbed on a single-atom vacancy obtained with the PBE functional (neglecting the vdW interaction).}
\end{center}
\label{tab:geo-vdW}
\end{table}

\begin{figure}[bhtp]
\begin{center}
\includegraphics[width=0.5\textwidth]{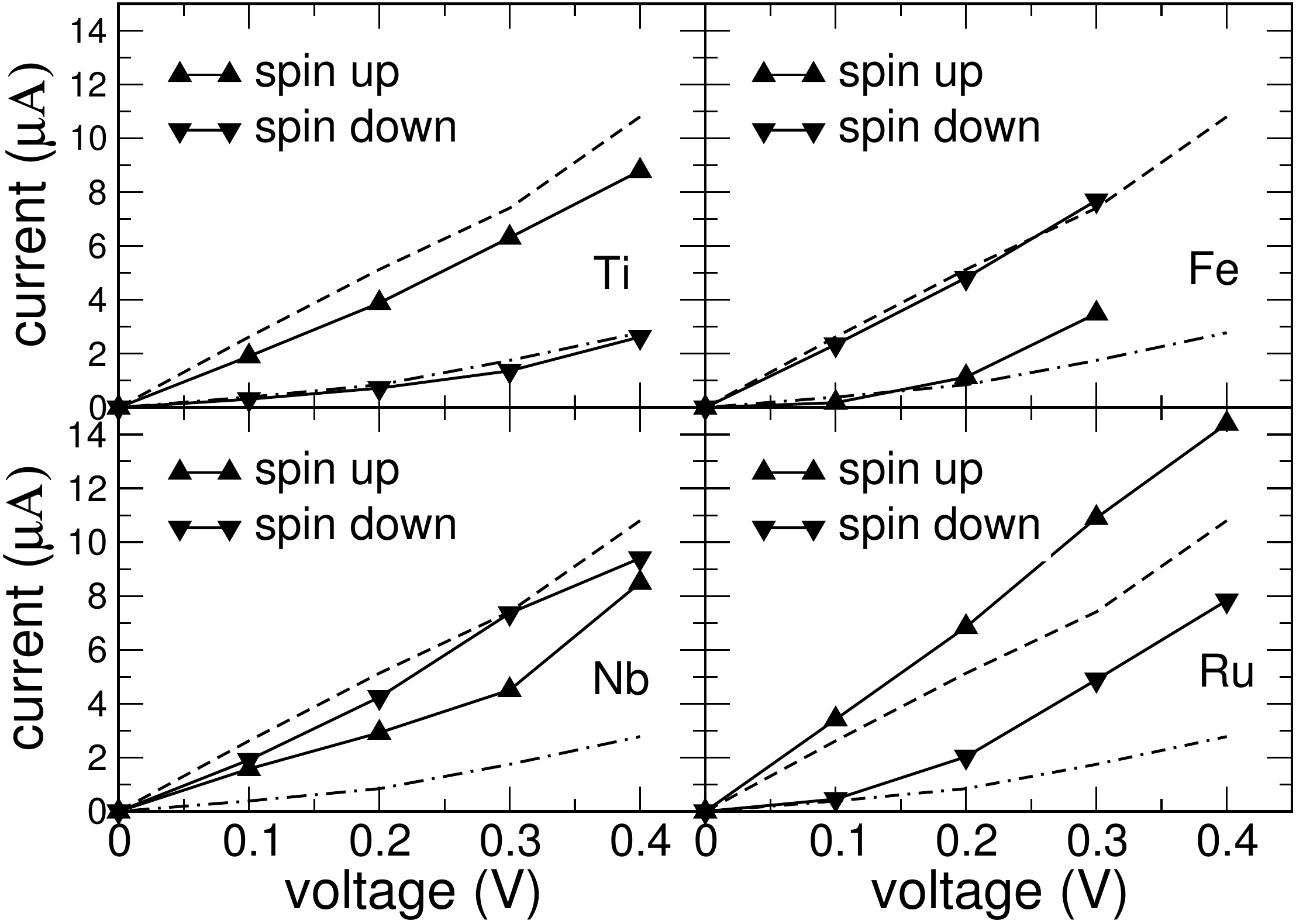} 
\caption{Spin-resolved current (in $\mu$A) as function of source-drain voltage for the combined system of a metal porphyrin adsorbed on a single-atom vacancy in a zig-zag edge graphene nanoribbon, whose geometry was optimized without van der Waals forces. 
The upper row shows the currents for Ti-porphyrin (left) and Fe-porphyrin (right), the lower row for Nb-porphyrin (left) and Ru-porphyrin (right).
The lines with triangular symbols show the currents in the spin channels. 
For comparison, the spin currents of the clean GNR with single-atom vacancy are shown as dashed (spin-up) and dash-dotted (spin-down) lines in all subfigures. 
}
\label{fig:transport-vdW}
\end{center}
\end{figure}

Interestingly, while the incorporation of most metal atom dopants maintain a planar structure, the elements 
Ca, Nb, Sc, Y and Zr have their most stable positions outside the atomic plane of porphyrin. 
The calculated magnetic moment values can be first understood by considering an ion in the square planar crystal field. 
For the case of the late transition metals, such as Co and Rh, the bond distances of the metal centre to the nitrogen atoms in the ring are so small that the $d$ electrons remaining at the metal prefer to form a low-spin state. Similarly, for early transition metals, a high-spin (such as Cr, Mo, Nb, Tc, and V) or intermediate state (such as Fe, Mn, and Ru) are preferred. For the case of Ni, the spin suppression is due to the strong structural distortion leading to the further splitting of the $d_{xz}$ and $d_{yz}$ orbitals.

Out of the listed porphyrins, four examples have been selected in order to study the interaction of metal porphyrins with graphene. 
 As examples of early transition metals with valency four or five, we selected Ti-porphyrin and Nb-porphyrin. 
 As representatives for the late transition metals, we took the isovalent species Fe and Ru. 
 
 \subsection{Effect of the van der Waals interaction on transport}
 
 While the van der Waals (vdW) interaction is clearly important for the adsorption energy of the porphyrin molecules on the graphene nanoribbon (GNR), its role for the electrical transport is mostly an indirect one, mediated by the changes of the adsorption geometry due to the vdW interaction. 
 In order to assess the sensitivity of the transport results to geometrical details, we carried out the transport calculations not only for the geometries relaxed including vdW forces, but in addition for relaxed geometries where vdW forces had been discarded. 
 It is observed that, without vdW, the interaction between the porphyrin and the GNR with the vacancy is less attractive. The organic side groups of the molecules and the GNR repel each other. While the metal centre is still bonded to the carbon atoms next to the vacancy with nearly the same bond lengths, the adsorption height $h$ of the porphyrin ring relative to the plane of the GNR is 0.2 to 0.3 {\AA} larger than in the case of vdW interaction included (see Table \ref{tab:geo-vdW}). The C atoms of the GNR that bond to the metal are lifted above the plane of the GNR. In the adsorption geometry optimized without the vdW interaction, the adsorption causes a more pronounced distortion of the GNR, whereas the GNR remains almost planar if the vdW interaction is taken into account.
 
 The transport calculation shows that the deviations from planarity have only a minor effect on the conductance of the GNR. 
 We attribute this finding to electronic states near the edges of the GNR that are almost unaffected by the presence of the molecule and the distortion of the central part of the GNR.
 The spin-polarized currents for the geometries optimized without vdW, displayed in Fg.~\ref{fig:transport-vdW}, are very similar to the currents in Fig. 5, where geometries optimized with vdW forces have been used. We conclude that the transport results are robust against uncertainties in the atomic geometries. The differences between the various adsorbed porphyrins are mainly due to charge transfer between the porphyrin and the GNR and, as a consequence,  different positioning of the Fermi level in each adsorption system.

\end{document}